\PassOptionsToPackage{obeyspaces}{url}
\documentclass[sigconf,screen]{acmart}

\usepackage[utf8]{inputenc}
\usepackage[T1]{fontenc}
\usepackage{microtype}

\usepackage{bm}
\usepackage{soul}
\usepackage{graphicx}
\usepackage{amsmath,amsfonts}
\usepackage{booktabs}
\usepackage{multirow}
\usepackage{makecell}
\usepackage{textcomp}
\usepackage{array}
\usepackage{textcomp}
\usepackage[font=footnotesize]{subfig}
\usepackage{multirow}
\usepackage{float}
\usepackage{xspace}
\usepackage[export]{adjustbox}
\usepackage{nicefrac}
\usepackage{colortbl}
\usepackage{tablefootnote}
\usepackage[referable]{threeparttablex}
\usepackage{comment}
\usepackage{listings}
\usepackage{mathtools}
\usepackage{enumitem}

\usepackage{wrapfig}
\BeforeBeginEnvironment{wrapfigure}{\setlength{\intextsep}{2pt}}

\makeatletter
\@ifpackageloaded{icml\the\year}{
\usepackage{enumitem}
\setlist{nosep}}
{\usepackage[linesnumbered,ruled,vlined]{algorithm2e}
\SetKwRepeat{Do}{do}{while}
\DontPrintSemicolon

\SetCommentSty{commentsty}
}

\@ifclassloaded{IEEEtran}{
\usepackage[numbers,sort,compress]{natbib}

}{}

\@ifclassloaded{acmart}{
\def\authornotetext#1{
	\g@addto@macro\@authornotes{%
	\stepcounter{footnote}\footnotetext{#1}}%
}}{
\usepackage[usenames,svgnames]{xcolor}
\usepackage{hyperref}
\hypersetup{
    colorlinks = true,
    breaklinks = true,
    bookmarksdepth = section,
    citecolor = DarkGreen,
    linkcolor = DarkRed,
    urlcolor = DarkBlue,
}}

\@ifclassloaded{llncs}{
\usepackage[square,comma,numbers,sort&compress]{natbib}
\bibliographystyle{splncsnat}

}{
\usepackage{amsthm}

\theoremstyle{remark}

\theoremstyle{definition}

}
\makeatother

\usepackage{pifont}

\definecolor{sblue}{HTML}{20C1EE}
\definecolor{sred}{HTML}{E54F44}
\definecolor{spink}{HTML}{DE47A2}
\definecolor{sgreen}{HTML}{1AAF54}
\definecolor{spurple}{HTML}{7869BE}
\definecolor{syellow}{HTML}{EDB429}

\DeclarePairedDelimiterX{\infdivx}[2]{(}{)}{%
	#1\;\delimsize\|\;#2%
}

\DeclareMathAlphabet{\mathsfit}{\encodingdefault}{\sfdefault}{m}{sl}
\SetMathAlphabet{\mathsfit}{bold}{\encodingdefault}{\sfdefault}{bx}{n}


\setcopyright{acmlicensed}
\copyrightyear{2021}
\acmYear{2021}
\acmConference[MM '21]{Proceedings of the 29th ACM International Conference on Multimedia}{October 20--24, 2021}{Virtual Event, China}
\acmBooktitle{Proceedings of the 29th ACM International Conference on Multimedia (MM '21), October 20--24, 2021, Virtual Event, China}
\acmPrice{15.00}
\acmDOI{10.1145/3474085.3475259}
\acmISBN{978-1-4503-8651-7/21/10}

\author[Jinghao Zhang, Yanqiao Zhu, Qiang Liu, Shu Wu, Shuhui Wang, and Liang Wang]{Jinghao Zhang$^{1,3,}$*, Yanqiao Zhu$^{1,3,}$*, Qiang Liu$^{1,3,\dagger}$, Shu Wu$^{1,3,4}$, Shuhui Wang$^{2}$, and Liang Wang$^{1,3}$}
\authornotetext{The first two authors made equal contribution to this work.}
\authornotetext{To whom correspondence should be addressed.}
\affiliation{
	\institution{$^1$Center for Research on Intelligent Perception and Computing, Institute of Automation, Chinese Academy of Sciences}
	\institution{$^2$Key Laboratory of Intelligent Information Processing, Institute of Computing Technology, Chinese Academy of Sciences}
	\institution{$^3$School of Artificial Intelligence, University of Chinese Academy of Sciences}
	\institution{$^4$Artificial Intelligence Research, Chinese Academy of Sciences}
	\country{}
}
\email{{jinghao.zhang,yanqiao.zhu}@cripac.ia.ac.cn, shuhui.wang@vipl.ict.ac.cn, {qiang.liu,shu.wu,wangliang}@nlpr.ia.ac.cn}
\def\authors{Jinghao Zhang, Yanqiao Zhu, Qiang Liu, Shu Wu, Shuhui Wang, and Liang Wang}

\begin{document}
\fancyhead{}

\newcommand{\themodel}{LATTICE\xspace}

\title{Mining Latent Structures for Multimedia Recommendation}

\begin{abstract}
Multimedia content is of predominance in the modern Web era. Investigating how users interact with multimodal items is a continuing concern within the rapid development of recommender systems.
The majority of previous work focuses on modeling user-item interactions with multimodal features included as side information. However, this scheme is not well-designed for multimedia recommendation. Specifically, only \emph{collaborative} item-item relationships are implicitly modeled through high-order item-user-item relations. Considering that items are associated with rich contents in multiple modalities, we argue that the latent \emph{semantic} item-item structures underlying these multimodal contents could be beneficial for learning better item representations and further boosting recommendation.
To this end, we propose a \underline{LAT}ent s\underline{T}ructure mining method for mult\underline{I}modal re\underline{C}omm\underline{E}ndation, which we term \themodel for brevity. To be specific, in the proposed \themodel model, we devise a novel modality-aware structure learning layer, which learns item-item structures for each modality and aggregates multiple modalities to obtain latent item graphs.
Based on the learned latent graphs, we perform graph convolutions to explicitly inject high-order item affinities into item representations. These enriched item representations can then be plugged into existing collaborative filtering methods to make more accurate recommendations.
Extensive experiments on three real-world datasets demonstrate the superiority of our method over state-of-the-art multimedia recommendation methods and validate the efficacy of mining latent item-item relationships from multimodal features.
\end{abstract}


\maketitle

\newcommand{\inlinegraphics}[1]{(\raisebox{-.1\height}{\includegraphics[width=0.9em]{#1}})}

\section{Introduction}

\begin{figure}
	\centering
	\includegraphics[width=\linewidth]{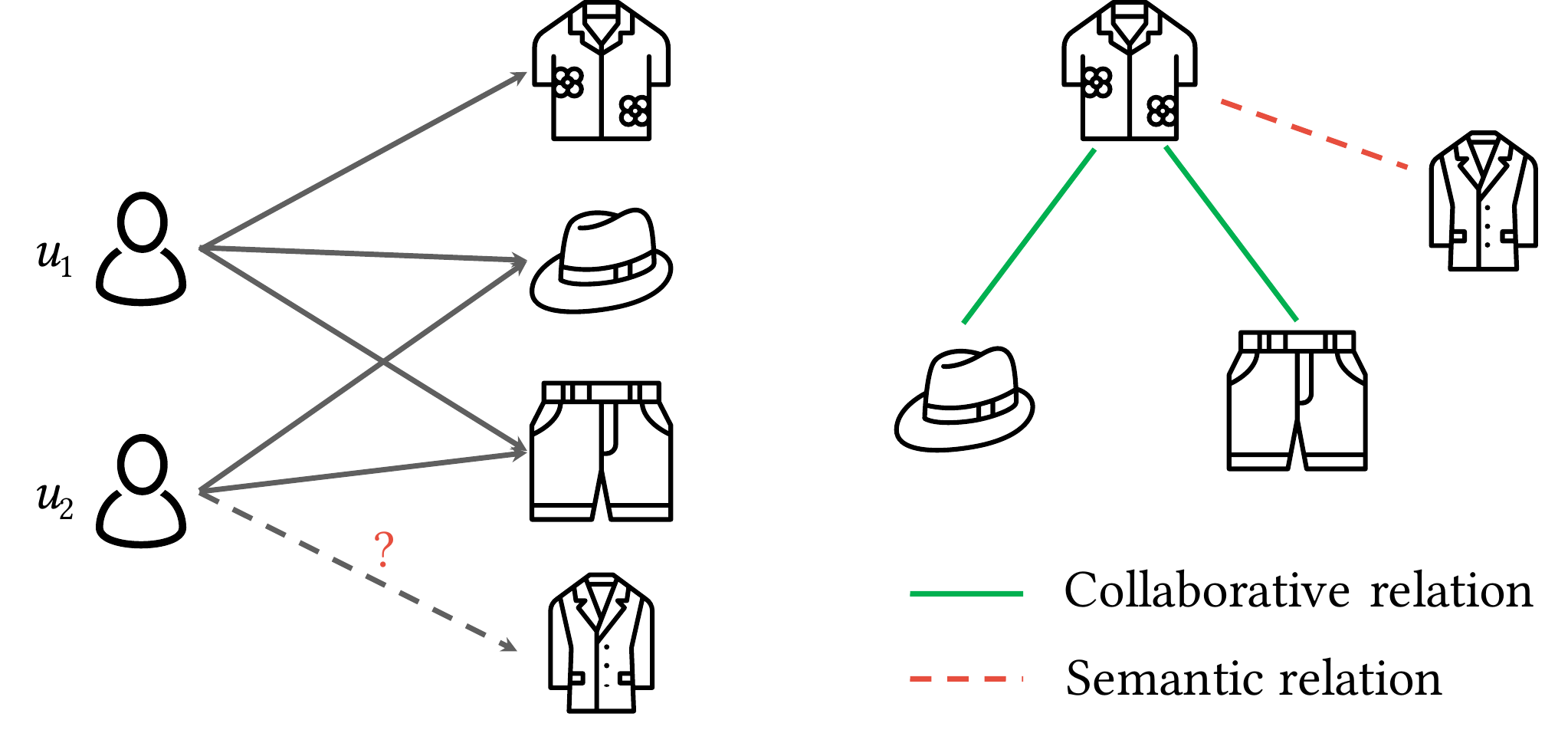}
	\caption{A toy example of recommendation with two types of item relations. In this paper, we argue that {\color[rgb]{0.89803922,0.30980392,0.26666667} semantic structures} mined from multimodal features are helpful for comprehensively discovering candidate items supplementary to {\color[rgb]{0.09803922,0.68627451,0.32941176} collaborative signals} in traditional work.}
	\label{fig:toy-example}
\end{figure}

\begin{figure*}
	\centering
    \includegraphics[width=\linewidth]{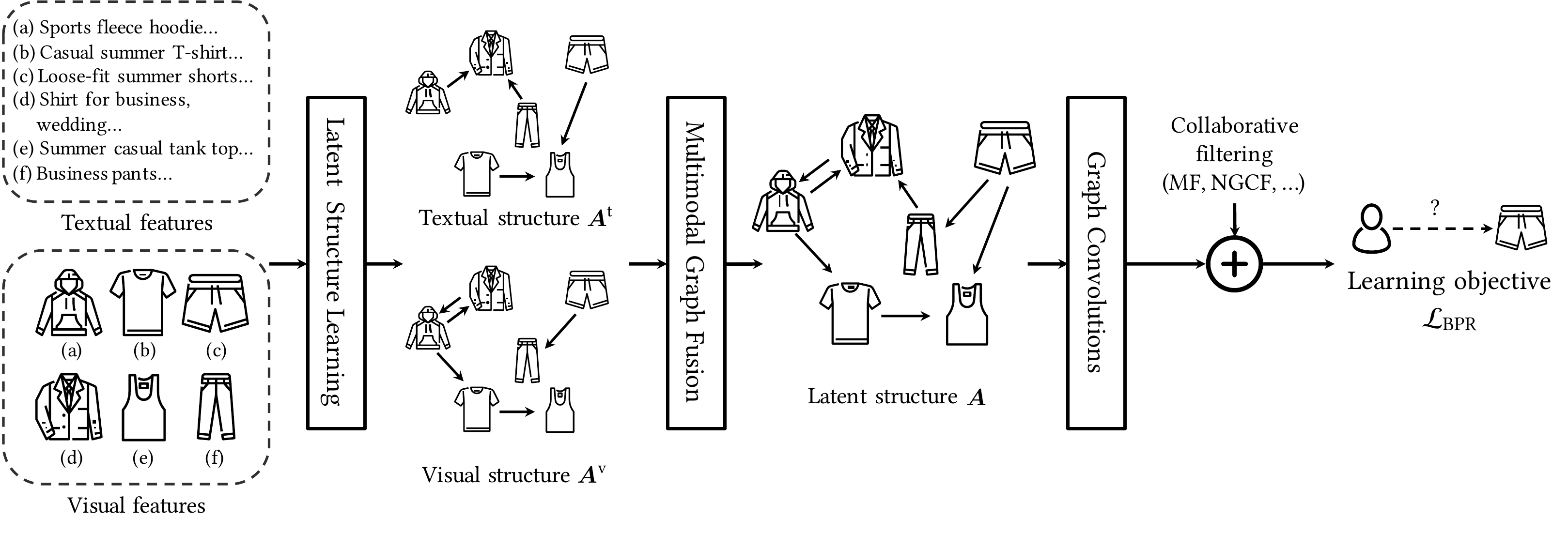}
    \caption{The overall framework of our proposed \themodel model. We first learn modality-aware item graphs and aggregate multiple modalities in an adaptive manner. Based on the mined latent graphs, we conduct graph convolutions to inject high-order item relationships into item embeddings, which are then combined with downstream CF methods to make recommendations.}
    \label{fig:model}
\end{figure*}

With the rapid development of Internet, recommender systems have become an indispensable tool to help users find relevant information. Nowadays, users are easily accessible to large amounts of online information represented in multiple modalities, including images, texts, videos, etc. Recent years have witnessed growing research interests on multimedia recommendation, which aims to predict whether a user will interact with an item with multimodal contents. It has been successfully applied to many online applications, such as e-commerce, instant video platforms, and social media platforms.

Focusing on exploiting abundant user-item interactions, collaborative filtering (CF) serves as a foundation of personalized recommender systems, which encodes users and items into low-dimensional dense vectors and makes recommendations based on these embeddings \cite{Aggarwal:2016dl,He:2017jw,Su:2009cl}.
Traditional work on multimedia recommendation, e.g., VBPR \cite{He:2016ww}, DeepStyle \cite{Liu:2017ij}, and ACF \cite{Chen:2017jj}, extends the vanilla CF framework by incorporating multimodal contents as side information in addition to ID embeddings of items. However, as these methods only model direct user-item interactions, their expressiveness is confined.

Inspired by the recent surge of graph neural networks \cite{Kipf:2017tc,Velickovic:2018we,Hu:2019vq}, \citet{Wang:2019er} propose to model user-item relationships as bipartite graphs and inject high-order interactions into the embedding process to learn better representations. These graph-based recommender systems \cite{He:2020gd,Wu:2019ke,Yu:2020dn} achieve great success and obtain state-of-the-art performance.
Recently, many attempts have been made to integrate multimodal contents into graph-based recommendation systems. MMGCN \cite{Wei:2019hn} constructs modality-specific user-item interaction graphs to model user preferences specific to each modality. Following MMGCN, GRCN \cite{Wei:2020ko} utilizes multimodal features to refine user-item interaction graphs by identifying false-positive feedbacks and prunes the corresponding noisy edges.

Despite their effectiveness, previous attempts fail to explicitly model item relationships, which have been proved to be important in recommender systems \cite{Sarwar:2001kx}.
Specifically, the majority of previous work concentrates on modeling user-item interactions by constructing better interaction graphs or designing sophisticated user-item aggregation strategies, following the traditional CF paradigm. Therefore, only \emph{collaborative} item relationships are implicitly discovered through modeling high-order item-user-item co-occurrences, which potentially leads to a gap to the genuine item-item relations that carry \emph{semantic} relationships.
Taking Figure \ref{fig:toy-example} as an example, existing methods will recommend the shirt \inlinegraphics{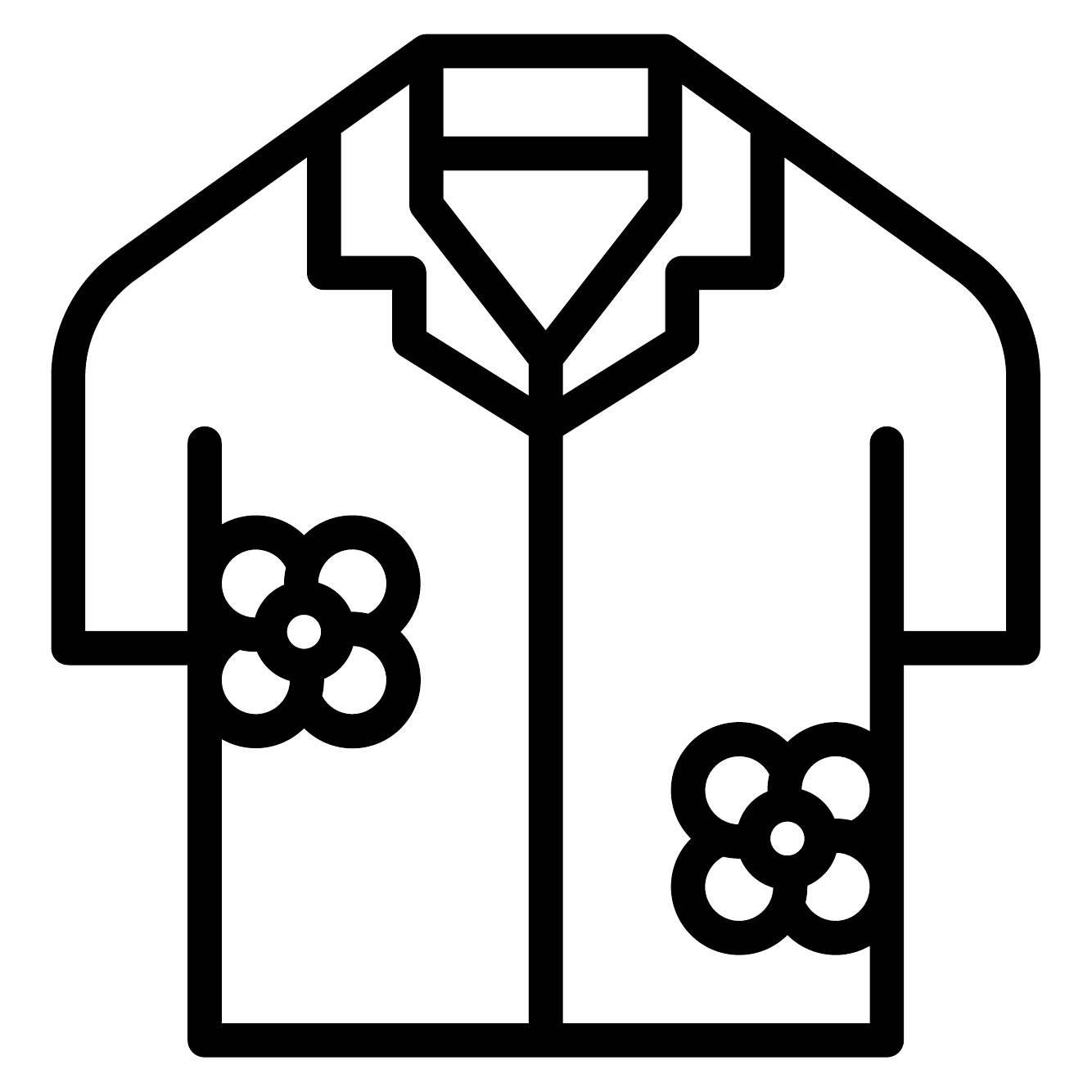} for \(u_2\) according to collaborative relationships, since shirts \inlinegraphics{figures/icons/hawaiian-shirt.pdf}, hats \inlinegraphics{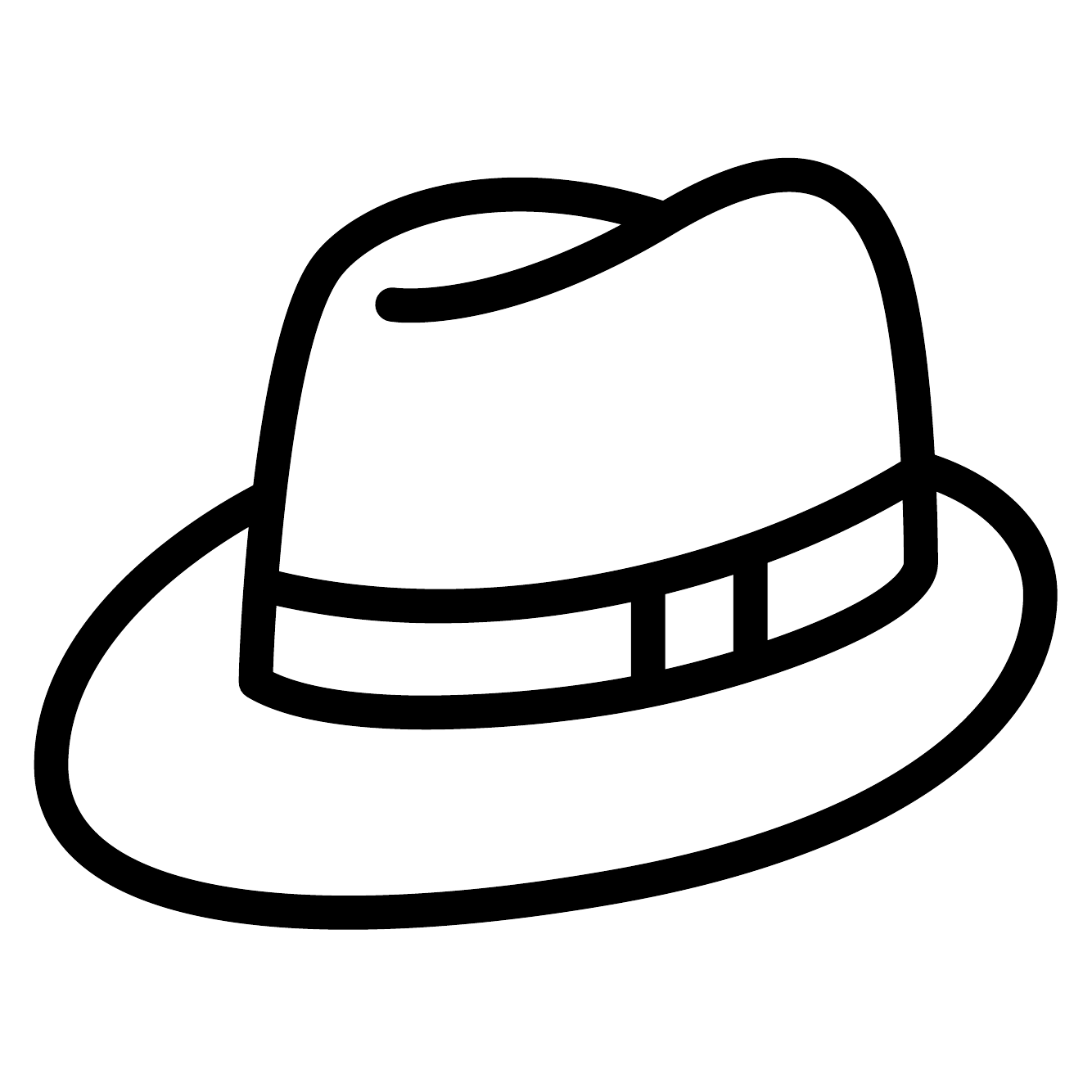}, and pants \inlinegraphics{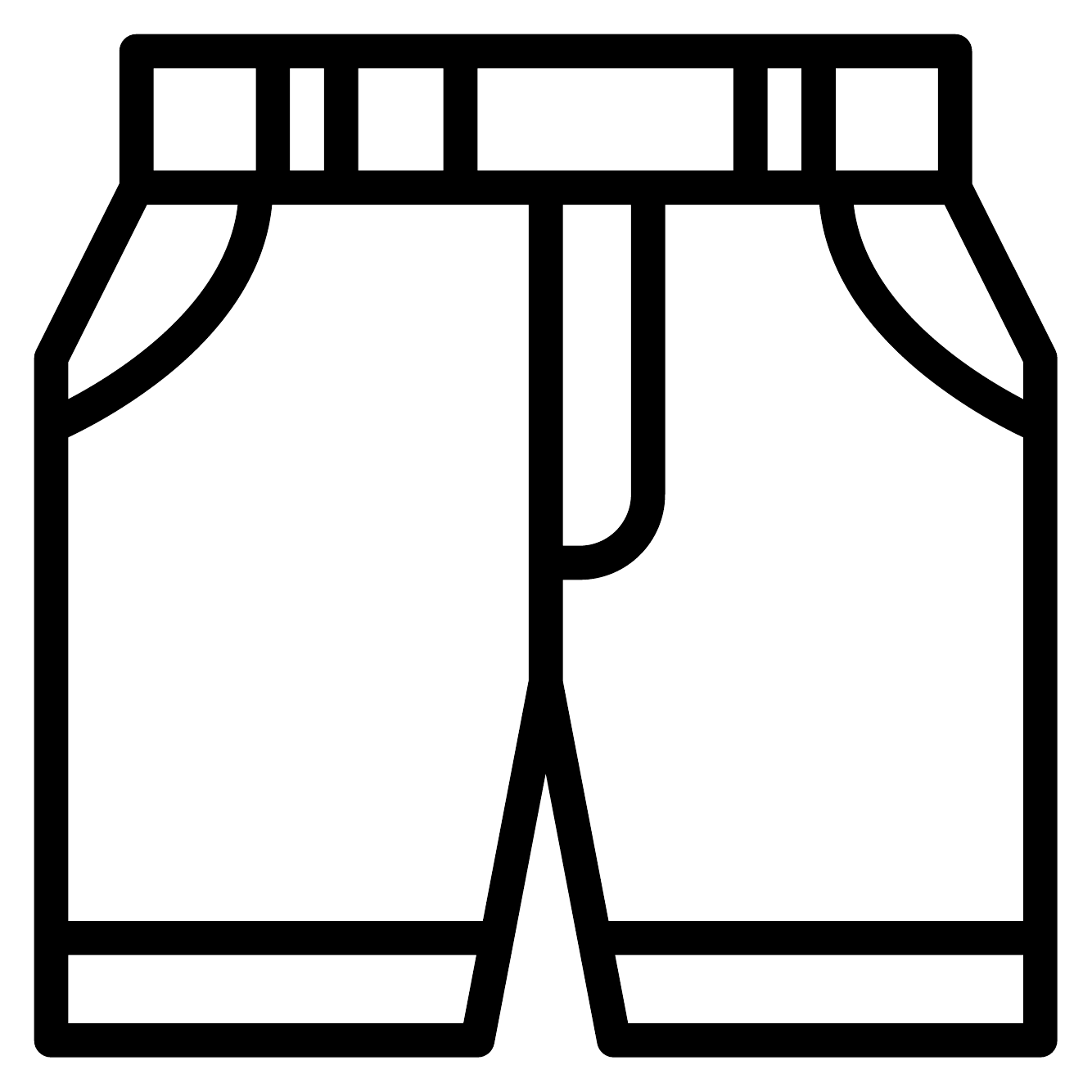} all interacted with \(u_1\). However, previous work may not be able to recommend coats \inlinegraphics{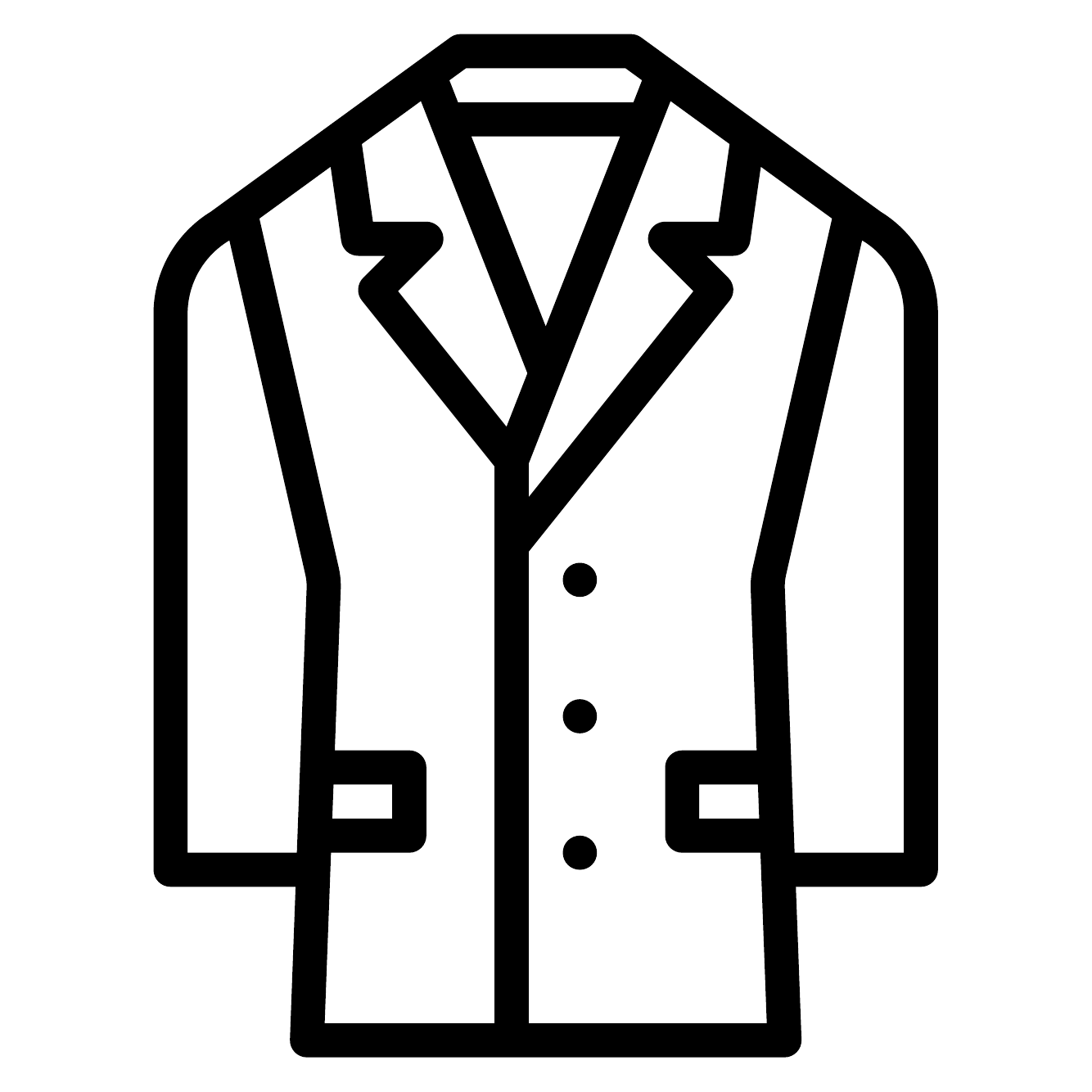} to \(u_2\), which are visually similar to shirts.
Considering that items are associated with rich multimodal features, the latent connections underlying multimodal contents would facilitate learning better item representations and assist the recommender models to comprehensively discover candidate items.

Towards this end, we propose a novel \underline{LAT}ent s\underline{T}ructure mining scheme for mult\underline{I}modal re\underline{C}omm\underline{E}ndation, \themodel for brevity.
As shown in Figure \ref{fig:model}, the proposed \themodel consists of three key components. Firstly, we develop a novel modality-aware structure learning layer, which learns modality-aware item structures from multimodal features and aggregates modality-aware item graphs to construct latent multimodal item graphs. After that, we perform graph convolutions on the learned latent graphs to explicitly consider item relationships.
The resulting item representations are thus infused with high-order item relationships, which will be added into the output item embeddings of CF models.
Please kindly note that distinct from previous work that leverages raw features of multimodal contents as side information in addition to ID embeddings, in our work, multimodal features are only used to learn graph structures, and graph convolutions are employed on ID embeddings to directly model \emph{item-item affinities}.

The \themodel model enjoys two additional benefits.
Firstly, previous work that leverages multimodal features to model user-item interactions faces the cold-start problem, when limited user-item interactions are provided. Our work, on the contrary, mines latent item graph structures from multimodal features. Even with limited interactions, items will get similar feedbacks from relevant neighbors through neighborhood aggregation.
Secondly, unlike previous attempts which design sophisticated user-item aggregation strategies, \themodel is agnostic to downstream CF methods. Therefore, it could be served as a play-and-plug module for existing recommender models.

In summary, the main contribution of this work is threefold.
\begin{itemize}
	\item We highlight the importance of explicitly exploiting item relationships in multimedia recommendation, which are helpful for discovering comprehensive candidate items.
	\item We propose a novel framework for multimedia recommendation to mine latent relationships beneath multimodal features, which supplement the collaborative signals modeled by traditional CF methods.
	\item We perform extensive experiments on three public datasets to validate the effectiveness of our proposed model.
\end{itemize}
To foster reproducible research, our code is made publicly available at \url{https://github.com/CRIPAC-DIG/LATTICE}.

\section{The Proposed Method}

In this section, we first formulate the problem and introduce our model in detail. As illustrated in Figure \ref{fig:model}, there are three main components in our proposed framework:
(1) a modality-aware graph structure learning layer that learns item graph structures from multimodal features and fuses multimodal graphs,
(2) graph convolutional layers that learn the embeddings by injecting item-item affinities,
and (3) downstream CF methods. 

\subsection{Preliminaries}
Let $\mathcal{U}$, $\mathcal{I}$ denote the set of users and items, respectively.  Each user $u \in \mathcal{U}$ is associated with a set of items $\mathcal{I}^u$ with positive feedbacks which indicate the preference score $y_{ui}=1$ for $i \in \mathcal{I}^u$. $\bm x_u, \bm x_i \in \mathbb{R}^{d}$ is the input ID embedding of $u$ and $i$, respectively, where $d$ is the embedding dimension. Besides user-item interactions, multimodal features are offered as content information of items. We denote the modality features of item $i$ as $\bm{e}_i^m \in \mathbb{R}^{d_m}$, where $d_m$ denotes the dimension of the features, $m \in \mathcal{M}$ is the modality, and $\mathcal{M}$ is the set of modalities.
The purpose of multimodal recommendation is to accurately predict users' preferences by ranking items for each user according to predicted preference scores $\hat{y}_{u i}$.
In this paper, we consider visual and textual modalities denoted by $\mathcal{M} = \{ \text{v}, \text{t} \}$. Please kindly note that our method is not fixed to the two modalities and multiple modalities can be involved.

\subsection{Modality-aware Latent Structure Learning}
Multimodal features provide rich and meaningful content information of items, while existing methods only utilize multimodal features as side information for each item, ignoring the important \emph{semantic} relationships of items underlying features. In this section, we introduce how to discover the underlying latent graph structure of item graphs in order to learn better item representations.
To be specific, we first construct initial $k$-nearest-neighbor ($k$NN) modality-aware item graphs by utilizing raw multimodal features. After that, we learn the latent graph structures from transformed, high-level features based on the initial graph. Finally, we integrate the latent graphs from multiple modalities in an adaptive way. 

\subsubsection{Constructing initial $k$NN modality-aware graphs}
We first construct initial $k$NN modality-aware graph $\bm{S}^m$ by using raw features for each modality $m$. Based on the hypothesis that similar items are more likely to interact than dissimilar items \cite{McPherson:2003dp}, we quantify the \emph{semantic} relationship between two items by their similarity.
Common options for node similarity measurement include cosine similarity \cite{Wang:2020bs}, kernel-based functions \cite{Li:2018wu}, and attention mechanisms \cite{Chen:2020wu}. Our method is agnostic to similarity measurements, and we opt to the simple and parameter-free cosine similarity in this paper. The similarity matrix $\bm{S}^m \in \mathbb{R}^{N \times N}$ is computed by
\begin{equation}
    \bm{S}_{i j}^m = \frac{{(\bm{e}_i^m)}^\top \bm{e}_j^m}{ \| \bm{e}_i^m \| \| \bm{e}_j^m \|}.
    \label{eq:sim}
\end{equation}
Typically, the graph adjacency matrix is supposed to be non-negative but $\bm{S}_{i j}$ ranges between $[-1, 1]$. Thus, we suppress its negative entries to zeros.
Moreover, common graph structures are much sparser other than a fully-connected graph, which is computationally demanding and might introduce noisy, unimportant edges \cite{Chen:2020wu}. For this purpose, we conduct $k$NN sparsification \cite{Chen:2009wc} on the dense graph: for each item $i$, we only keep edges with the top-\(k\) confidence scores:
\begin{equation}
    \widehat{\bm{S}}^m_{i j}=
    \begin{cases}
    \bm{S}^m_{i j}, \enspace & {\bm{S}}^m_{i j} \in \operatorname{top-}k({\bm{S}}^m_{i}), \\
    0, \enspace & \text{otherwise},
    \end{cases}
    \label{eq:topk}
\end{equation}
where $\widehat{\bm{S}}^m$ is the resulting sparsified, directed graph adjacency matrix. To alleviate the exploding or vanishing gradient problem \cite{Kipf:2017tc}, we normalize the adjacency matrix as:
\begin{equation}
    \widetilde{\bm{S}}^m = ({\bm{D}^m})^{-\frac{1}{2}} \widehat{\bm{S}}^m ({\bm{D}^m})^{-\frac{1}{2}},
    \label{eq:norm}
\end{equation}
where $\bm{D}^m \in \mathbb{R}^{N \times N}$ is the diagonal degree matrix of $\widehat{\bm{S}}^m$ and $\bm{D}_{ii}^m = \sum_{j}\widehat{\bm{S}}^m_{i j}$.

\subsubsection{Learning latent structures}
Although we have obtained the modality-aware initial graph structures $\widetilde{\bm{S}}^m$ by utilizing raw multimodal features, they may not be ideal for the recommendation task. This is because the raw multimodal features are often noisy or even incomplete due to the inevitably error-prone data measurement or collection.  Additionally, initial graphs are constructed from the original multimodal features, which may not reflect the genuine graph structures after feature extraction and transformation. To this end, we propose to dynamically learn the graph structures by transformed, high-level multimodal features and combine learned structures with initial structures.

Firstly, we transform raw modality features into high-level features $\widetilde{\bm{e}}_{i}^m$:
\begin{equation}
    \widetilde{\bm{e}}_{i}^m = \bm{W}_m {\bm{e}}_{i}^m + \bm{b}_m,
\end{equation}
where ${\bm{W}}_m \in \mathbb{R}^{d' \times d_m}$ and ${\bm{b}}_m \in \mathbb{R}^{d'}$ denote the trainable transformation matrix and bias vector, respectively. $d'$ is the dimension of high-level feature vector $\widetilde{\bm{e}}_{i}^m$. We then dynamically infer the graph structures utilizing $\widetilde{\bm{e}}_{i}^m$, repeat the graph learning process described in Eqs. (\ref{eq:sim}, \ref{eq:topk}, \ref{eq:norm}) and obtain the adjacency matrix $\widetilde{\bm{A}}^m$.


Although the initial graph could be noisy, it typically still carries rich and useful information regarding item graph structures. Also, drastic change of adjacency matrix will lead to unstable training. To keep rich information of initial item graph and stabilize the training process, we add a skip connection that combines the learned graph with the initial graph:
\begin{equation}
    \bm{A}^m =  \lambda \widetilde{\bm{S}}^m + (1 - \lambda) \widetilde{\bm{A}}^m \ ,
\end{equation}
where $\lambda \in (0, 1)$ is the coefficient of skip connection that controls the amount of information from the initial structure. The obtained $\bm{A}^m$ is the final graph adjacency matrix representing latent structures for modality \(m\).
It is worth mentioning that both $\widetilde{\bm{S}}^m$ and $\widetilde{\bm{A}}^m$ are sparsified and normalized matrices, thus the final adjacency matrix $\bm{A}^m$ is also sparsified and normalized, which is computationally efficient and stabilizes gradients.

\subsubsection{Aggregating multimodal latent graphs}
After we obtained modality-aware adjacency matrix $\bm{A}^m$ for each modality $m \in \mathcal{M}$, in this section, we explore how to fuse different modalities to compute the final latent structures.
In multimedia recommendation, users usually focus on different modalities in different scenarios. For example, one may pay more attention to visual modality when selecting clothes, while focusing more on textual information when picking books, we thus introduce learnable weights to assign different importance scores to modality-specific graphs in an adaptive way:
\begin{equation}
	{\bm{A}} = \sum_{m=0}^{|\mathcal{M}|} \alpha_{m} \bm{A}^{m},
\end{equation}
where \(\alpha_m\) is the importance score of modality $m$ and ${\bm{A}} \in \mathbb{R}^{N \times N}$ is the graph that represents multimodal item relationships. We  apply the softmax function to keep the adjacency matrix $\bm{{A}}$ normalized, such that \(\sum_{m=0}^{|\mathcal{M}|} \alpha_{m} = 1\).

\subsection{Graph Convolutions}
\label{sec:graph-conv}
After obtained the latent structures, we perform graph convolution operations to learn better item representations by injecting item-item affinities into the embedding process. Graph convolutions can be treated as message propagating and aggregation. Through propagating the item representations from its neighbors, one item can aggregate the information within the first-order neighborhood. Furthermore, by stacking multiple graph convolutional layers, the high-order item-item relationships can be captured.

Following \citet{Wu:2019vz,He:2020gd}, we employ simple message passing and aggregation without feature transformation and non-linear activations which is effective and computationally efficient. In the $l$-th layer, the message passing and aggregation could be formulated as:
\begin{equation}
    \bm{h}_{i}^{(l)} = \sum_{j \in \mathcal{N}(i)} {\bm{A}}_{ij} \bm{h}_{j}^{(l-1)},
\end{equation}
where $\mathcal{N}(i)$ is the neighbor items and $\bm{h}_{i}^{(l)} \in \mathbb{R}^{d}$ is the $l$-th layer item representation of item $i$. We set the input item representation $\bm{h}_{i}^{(0)}$ as its corresponding ID embedding vector $\bm{x}_{i}$. We utilize ID embeddings of items as input representations rather than multimodal features, since we employ graph convolutions in order to directly capture item-item affinities. After stacking $L$ layers, $\bm{h}_{i}^{(L)}$ encodes the high-order item-item relationships that are constructed by multimodal information and thus can benefit the downstream CF methods.

\subsection{Combining with Collaborative Filtering}
\label{CF}
Different from previous attempts which design sophisticated user-item aggregation strategies, \themodel learns item representations from multimodal features and then combine them with downstream CF methods that model user-item interactions. It is flexible and could be served as a play-and-plug module for any CF methods. 



We denote the output user and item embeddings from CF methods as $\widetilde{\bm{x}}_u, \widetilde{\bm{x}}_i \in \mathbb{R}^d$ and simply enhance item embeddings by adding normalized item embeddings $\bm{h}_{i}^{(L)}$ learned through item graph:
\begin{equation}
    \widehat{\bm{x}}_{i} = \widetilde{\bm{x}}_i + \frac{\bm{h}_{i}^{(L)}}{\|\bm{h}_{i}^{(L)}\|_2}.
    \label{eq:cf}
\end{equation}
We then compute the user-item preference score by taking inner product of user embeddings and enhanced item embeddings:
\begin{equation}
    \hat{y}_{u i} = \widetilde{\bm{x}}_u^\top \widehat{\bm{x}}_{i}.
\end{equation}

We conduct experiments in Section \ref{sec:rq2} on different downstream CF methods. The play-and-plug paradigm separates the usage of multimodal features with user-item interactions, thus alleviating the cold-start problem, where tailed items are only interacted with few users or even never interacted with users. We learn latent structures for items and the tailed items will get similar feedbacks from relevant neighbors through neighborhood aggregation. We conduct experiments in cold-start settings in Section \ref{sec:rq1} which proves the effectiveness of this play-and-plug paradigm.

\subsection{Optimization}
We adopt Bayesian Personalized Ranking (BPR) loss \cite{Rendle:2009wp} to compute the pair-wise ranking, which encourages the prediction of an observed entry to be higher than its unobserved counterparts:
\begin{equation}
    \mathcal{L}_{\text{BPR}}=-\sum_{u \in \mathcal{U}} \sum_{i \in \mathcal{I}_{u}} \sum_{j \notin \mathcal{I}_{u}} \ln \sigma\left(\hat{y}_{u i}-\hat{y}_{u j}\right),
\end{equation}
where $\mathcal{I}^{u}$ indicates the observed items associated with user $u$ and $(u, i, j)$ denotes the pairwise training triples where $i \in \mathcal{I}^u$ is the positive item and $j \notin \mathcal{I}^u$ is the negative item sampled from unobserved interactions. $\sigma(\cdot)$ is the sigmoid function.

\section{Experiments}
In this section, we conduct experiments on three widely used real-world datasets to answer the following research questions:
\begin{itemize}
	\item \textbf{RQ1:} How does our model perform compared with the state-of-the-art multimedia recommendation methods and other CF methods in both warm-start and cold-start settings?
	\item \textbf{RQ2:} How effective are the item graph structures learned from multimodal features?
	\item \textbf{RQ3:} How sensitive is our model under the perturbation of several key hyper-parameters?
\end{itemize}

\subsection{Experiments Settings}
\subsubsection{Datasets}

\begin{table}[t]
\begin{threeparttable}[t]
	\caption{Statistics of the datasets}
	\begin{tabular}{ccccc}
	\toprule
	Dataset\tnotex{tn:link} & \# Users & \# Items & \# Interactions & Density \\
	\midrule
	Clothing & 39,387   & 23,033   & 237,488         & 0.00026 \\
	Sports   & 35,598   & 18,357   & 256,308         & 0.00039 \\
	Baby     & 19,445   & 7,050    & 139,110         & 0.00101 \\
	\bottomrule
	\end{tabular}
	\label{tab:dataset}
	\begin{tablenotes}[flushleft]
	\footnotesize{
		\item[1]\label{tn:link} Datasets can be accessed at \url{http://jmcauley.ucsd.edu/data/amazon/links.html}.
	}
	\end{tablenotes}
\end{threeparttable}
\end{table}

\begin{table*}
	\begin{tabular}{cccccccccc}
	\toprule
	\multirow{2.5}{*}{Model} & \multicolumn{3}{c}{Clothing}       & \multicolumn{3}{c}{Sports}         & \multicolumn{3}{c}{Baby}           \\
	\cmidrule(lr){2-4} \cmidrule(lr){5-7} \cmidrule(lr){8-10}
	& R@20 & P@20 & NDCG@20 & R@20 & P@20 & NDCG@20 & R@20 & P@20 & NDCG@20 \\ \midrule
	MF       & 0.0191          & 0.0010          & 0.0088          & 0.0430          & 0.0023          & 0.0202          & 0.0440          & 0.0024          & 0.0200          \\
	NGCF     & 0.0387          & 0.0020          & 0.0168          & 0.0695          & 0.0037          & 0.0318          & 0.0591          & 0.0032          & 0.0261          \\
	LightGCN & 0.0470          & 0.0024          & 0.0215          & 0.0782          & 0.0042          & 0.0369          & 0.0698          & 0.0037          & 0.0319           \\ \midrule
	VBPR     & 0.0481          & 0.0024          & 0.0205          & 0.0582          & 0.0031          & 0.0265          & 0.0486          & 0.0026          & 0.0213          \\
	MMGCN    & 0.0501          & 0.0024          & 0.0221          & 0.0638          & 0.0034          & 0.0279          & 0.0640          & 0.0032          & 0.0284          \\
	GRCN    & \underline{0.0631}         &\underline{0.0032}          & \underline{0.0276}          & \underline{0.0833}          & \underline{0.0044}          & \underline{0.0377}          & \underline{0.0754}          & \underline{0.0040}          & \underline{0.0336}          \\ \midrule
	\rowcolor{gray!20}
	\themodel     & \textbf{0.0710} & \textbf{0.0036} & \textbf{0.0316} & \textbf{0.0915} & \textbf{0.0048} & \textbf{0.0424} & \textbf{0.0829} & \textbf{0.0044} & \textbf{0.0368} \\
	Improv.  & 12.5\%          & 12.2\%          & 14.6\%          & 9.8\%          & 8.7\%          & 12.5\%          & 9.9\%           & 9.2\%           & 9.5\%        \\
	\bottomrule
	\end{tabular}
	\caption{Performance comparison of our \themodel with different baselines in terms of Recall@20 (R@20), Precision@20 (P@20), and NDCG@20. The best performance is highlighted \textbf{in bold} and the second to best is highlighted by \underline{underlines}. Improv. indicates relative improvements over the best baseline in percentage. All improvements are significant with \(p\)-value \(\leq 0.05\).}
	\label{tab:experiments}
\end{table*}
	
We conduct experiments on three categories of widely used Amazon dataset introduced by \citet{McAuley:2015ip}: \textsf{(a) Clothing, Shoes and Jewelry}, \textsf{(b) Sports and Outdoors}, and \textsf{(c) Baby}, which we refer to as \textbf{Clothing}, \textbf{Sports}, and \textbf{Baby} in brief.
The statistics of these three datasets are summarized in Table \ref{tab:dataset}. The three datasets include both visual and textual modalities. We use the 4,096-dimensional visual features that have been extracted and published. For the textual modality, we extract textual embeddings by concatenating the title, descriptions, categories, and brand of each item and utilize sentence-transformers \cite{Reimers:2019iz} to obtain 1,024-dimensional sentence embeddings.

\subsubsection{Baselines}
To evaluate the effectiveness of our proposed model, we compare it with several state-of-the-art recommendation models. These baselines fall into two groups: CF methods (i.e. MF, NGCF, LightGCN) and deep content-aware recommendation models (i.e. VBPR, MMGCN, GRCN).
\begin{itemize}
	\item \textbf{MF} \cite{Rendle:2009wp} optimizes Matrix Factorization using the Bayesian personalized ranking (BPR) loss, which exploits the user-item direct interactions only as the target value of interaction function.
	\item \textbf{NGCF} \cite{Wang:2019er} explicitly models user-item interactions by a bipartite graph. By leveraging graph convolutional operations, it allows the embeddings of users and items interact with each other to harvest the collaborative signals as well as high-order connectivity signals.
	\item \textbf{LightGCN} \cite{He:2020gd} argues the unnecessarily complicated design of GCNs (i.e. feature transformation and nonlinear activation) for recommendation systems and proposes a light model which only consists of two essential components: light graph convolution and layer combination.
	\item \textbf{VBPR} \cite{He:2016ww}: Based upon the BPR model, it integrates the visual features and ID embeddings of each item as its representation and feed them into Matrix Factorization framework. In our experiments, we concatenate multi-modal features as the content information to predict the interactions between users and items.
	\item \textbf{MMGCN} \cite{Wei:2019hn} is one of the state-of-the-art multimodal recommendation methods, which constructs modal-specific graphs and refines modal-specific representations for users and items. Tt aggregates all model-specific representations to obtain the representations of users or items for prediction.
	\item \textbf{GRCN} \cite{Wei:2020ko} is also one of the state-of-the-arts multimodal recommendation methods. It refines user-item interaction graph by identifying the false-positive feedback and prunes the corresponding noisy edges in the interaction graph.
\end{itemize}

\subsubsection{Evaluation protocols.}
We conduct experiments in both warm-start and cold-start settings.

\textbf{Warm-start settings.}
For each dataset, we select 80\% of historical interactions of each user to constitute the training set, 10\% for validation set, and the remaining 10\% for testing set. For each observed user-item interaction, we treat it as a positive pair, and then conduct the negative sampling strategy to pair them with one negative item that the user does not interact before.

\textbf{Cold-start settings.}
We remove all user-item interaction pairs associated with a randomly selected 20\% item set from the training set. We further divide the half of the items (10\%) into the validation set and half (10\%) into the testing set. In other words, these items are entirely unseen in the training set.

We adopt three widely-used metrics to evaluate the performance of preference ranking: Recall@\(k\), NDCG@\(k\), and Precision@\(k\). By default, we set \(k = 20\) and report the averaged metrics for all users in the testing set.

\subsubsection{Implementation details.}
We implemente our method in PyTorch \cite{Paszke:2019vf} and set the embedding dimension $d$ fixed to 64 for all models to ensure fair comparison. We optimize all models with the Adam \cite{Kingma:2015us} optimizer, where the batch size is fixed at 1024. We use the Xavier initializer \cite{Glorot:2010uc} to initialize the model parameters. The optimal hyper-parameters are determined via grid search on the validation set: the learning rate is tuned amongst \(\{0.0001, 0.0005, 0.001,\) \(0.005\}\), the coefficient of $\ell_2$ normalization is searched in \(\{0, 10^{-5},\) \(10^{-4}, 10^{-3}\}\), and the dropout ratio in \(\{0.0, 0.1, \cdots, 0.8\}\). Besides, we stop training if Recall@20 on the validation set does not increase for 10 successive epochs to avoid overfitting.

\subsection{Performance Comparison (RQ1)}
\label{sec:rq1}

\begin{table*}
    \begin{tabular}{cccccccccc}
    \toprule
    \multirow{2.5}{*}{Model} & \multicolumn{3}{c}{Clothing}       & \multicolumn{3}{c}{Sports}         & \multicolumn{3}{c}{Baby}           \\ \cmidrule(lr){2-4} \cmidrule(lr){5-7} \cmidrule(lr){8-10}
    & R@20 & P@20 & NDCG@20 & R@20 & P@20 & NDCG@20 & R@20 & P@20 & NDCG@20 \\
    \midrule
    MF             & 0.0191    & 0.0010       & 0.0088  & 0.0430               & 0.0023               & 0.0202               & 0.0440               & 0.0024               & 0.0200               \\
    MF+feats       & 0.0456    & 0.0023       & 0.0197  & 0.0674               & 0.0036               & 0.0304               & 0.0701               & 0.0037               & 0.0306               \\
    \themodel/feats--MF       & 0.0519    & 0.0026       & 0.0224  & 0.0708               & 0.0038               & 0.0319                & 0.0729               & 0.0037               & 0.0326               \\
    \rowcolor{gray!20}
    \themodel--MF        & 0.0577    & 0.0029       & 0.0246  & 0.0753               & 0.0040               & 0.0336               & 0.0767               & 0.0040               & 0.0339               \\
    Improv.       & 26.5\%    & 25.9\%       & 24.7\%  & 11.7\%               & 11.4\%               & 10.7\%               & 9.4\%                & 9.4\%                & 10.6\%    \\
    \midrule
    NGCF           & 0.0387    & 0.0020       & 0.0168  & 0.0695               & 0.0037               & 0.0318               & 0.0591               & 0.0032               & 0.0261               \\
    NGCF+feats     & 0.0436    & 0.0022       & 0.0190  & 0.0748               & 0.0040               & 0.0344               & 0.0660               & 0.0035               & 0.0295               \\
    \themodel/feats--NGCF       & 0.0480    & 0.0024       & 0.0212  & 0.0849               & 0.0043               & 0.0374               & 0.0713               & 0.0037               & 0.0307               \\
    \rowcolor{gray!20}
    \themodel--NGCF      & 0.0488    & 0.0025       & 0.0216  & 0.0856               & 0.0045               & 0.0381               & 0.0727               & 0.0039               & 0.0313               \\
    Improv.       & 12.0\%    & 11.9\%       & 13.7\%  & 14.5\%               & 14.2\%               & 10.9\%               & 10.1\%               & 9.4\%                & 6.0\%               \\
    \midrule
    LightGCN       & 0.0470    & 0.0024       & 0.0215  & 0.0782               & 0.0042               & 0.0369               & 0.0698               & 0.0037               & 0.0319               \\
    LightGCN+feats & 0.0477    & 0.0024       & 0.0208  & 0.0754               & 0.0040               & 0.0350               & 0.0793               & 0.0042               & 0.0344               \\
    \themodel/feats--LightGCN       & 0.0643    & 0.0033       & 0.0288  & 0.0832               & 0.0044               & 0.0386               & 0.0756               & 0.0040               & 0.0335               \\
    \rowcolor{gray!20}
    \themodel--LightGCN  & 0.0710    & 0.0036       & 0.0316  & 0.0915               & 0.0048               & 0.0424               & 0.0836               & 0.0044               & 0.0373               \\
    Improv.       & 48.8\%    & 48.4\%       & 52.0\%  & 21.3\%               & 20.5\%               & 21.3\%               & 5.4\%                & 5.2\%                & 8.3\%                \\
    \bottomrule
    \end{tabular}
    \caption{Performance of our proposed \themodel on top of different downstream collaborative filtering (CF) methods. Improv. indicates relative improvements in percentage over the base CF model with multimodal features (CF+feats).}
    \label{tab:ablation}
\end{table*}

We start by comparing the performance of all methods, and then explore how the item graph structures learned from multimodal features alleviate the cold-start problem. In this subsection, we combine our method with LightGCN as downstream CF method, and will also conduct experiments of different CF methods in Section \ref{sec:rq2}.

\subsubsection{Overall performance}
\label{sec:overall-perf}
Table \ref{tab:experiments} reports the performance comparison results, from which we can observe:

\begin{itemize}
	\item Our method significantly outperforms both CF methods and content-aware methods, verifying the effectiveness of our method. Specifically, our method improves over the strongest baselines in terms of Recall@20 by 12.5\%, 9.8\%, and 9.9\% in Clothing, Sports, and Baby, respectively. This indicates our proposed method is well-designed for multimodal recommendation by discovering underlying item-item relationships from multimodal features.
	\item Compared with CF methods, content-aware methods yield better performance overall, which indicates that multimodal features provide rich content information about items, and can boost recommendation accuracies. GRCN outperforms other baselines in three datasets since it discovers and prunes false-positive edges in user-item interaction graphs. Despite the sophisticated designed mechanisms, GRCN is still suboptimal compared to \themodel, which verifies the importance of explicitly capturing item-item relationships.
	\item Additionally, existing content-aware recommendation models are highly dependent on the representativeness of multimodal features and thus obtain fluctuating performances over different datasets. For Clothing dataset where visual features are very important in revealing item attributes \cite{Liu:2017ij,He:2016ww}, VBPR, MMGCN, and GRCN outperform all CF methods, including the powerful LightGCN. For the other two datasets where multimodal features may not directly reveal item attributes, content-aware methods obtain relatively small improvements. The performances of VBPR and MMGCN are even inferior to CF method LightGCN. Different from existing content-aware methods, we discover the latent item relationships underlying multimodal features instead of directly using them as side information. The latent item relationships are less dependent on the representativeness of multimodal features, and thus we are able to obtain better performance.
\end{itemize}

\subsubsection{Performance in cold-start settings.}

\begin{figure}
	\centering
	\includegraphics[width=\linewidth]{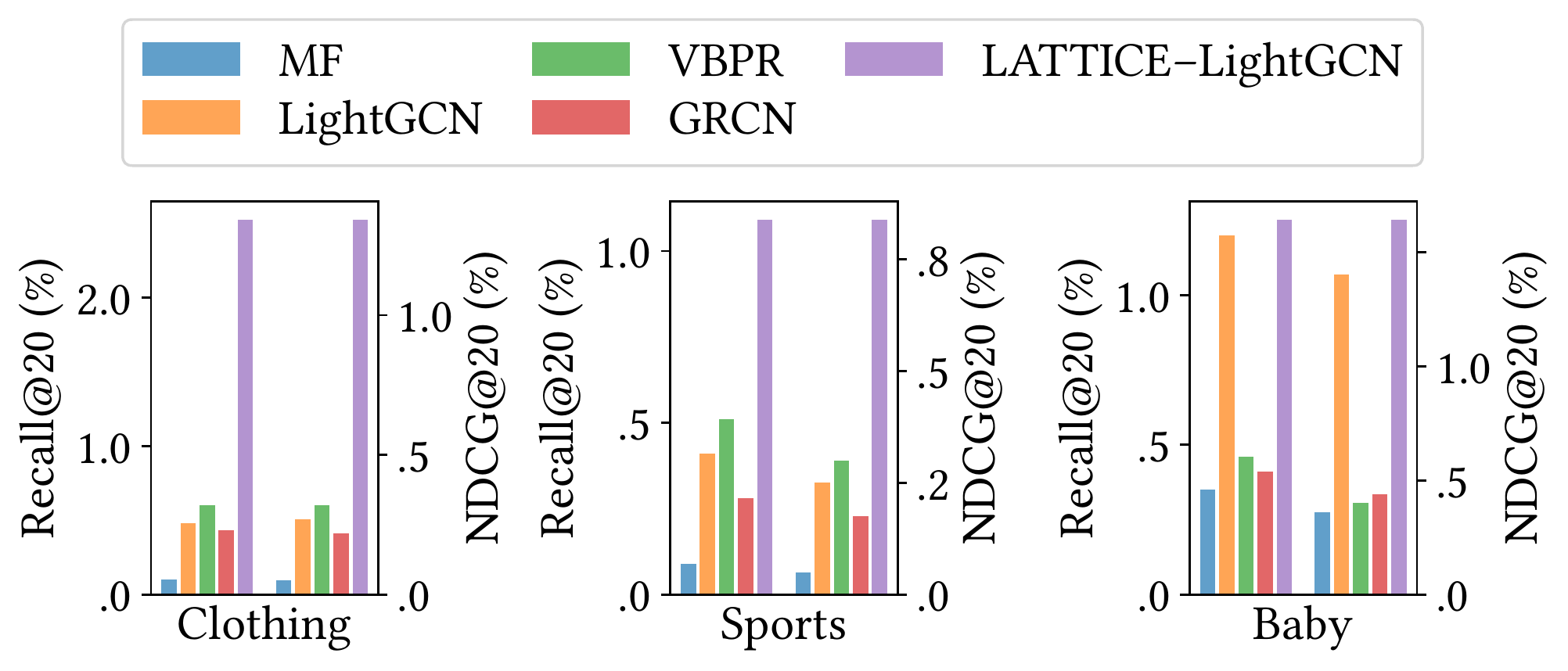}
	\caption{Performances of our method with different baselines in cold-start settings.}
	\label{fig:cold-start}
\end{figure}

The cold-start problem remains a prominent challenge in recommendation systems \cite{Schein:2002gw}. Multimodal features of items provide rich content information, which can be exploited to alleviate the  cold-start problem. We conduct cold start experiments and compare with representative baselines. Figure \ref{fig:cold-start} reports the results of performance comparison, from which we can observe:
\begin{itemize}
	\item \themodel can alleviate cold-start problem and outperforms all baselines in three datasets. We learn item graphs from multimodal features, along which cold-start items will get similar feedbacks from relevant neighbors through neighborhood aggregation.
	\item CF methods MF and LightGCN obtain poor performance under cold-start settings in general, primarily because they only leverage users’ feedbacks to predict the interactions between users and items. Although these methods may work well for items with sufficient feedbacks, they cannot help in cold-start settings, since no user-item interaction is available to update the representations of cold-start items.
	\item Content-aware model VBPR outperforms CF methods in general, which indicates the content information provided by multimodal features benefits recommendation for cold-start items. In particular, content information can help bridge the gap between the existing items to cold-start items. However, some graph-based content-aware methods such as GRCN, although performs well in warm-start settings, obtains poor performance in cold-start settings. GRCN utilizes multimodal features on user-item interaction bipartite graphs, which is also heavily dependent on user-item interactions. For cold-start items, they never interacted with users and become isolated nodes in the user-item graphs, leading to inferior performance.
\end{itemize}

\subsection{Ablation Studies (RQ2)}
\label{sec:rq2}

In this subsection, we combine \themodel with three common-used CF methods, i.e. MF, NGCF, and LightGCN to validate the effectiveness of our proposed method. For each CF method, we have three other variants: the first one is combined with our original method, employs graph convolutions on ID embeddings as described in Section \ref{sec:graph-conv}, named \textbf{\themodel--CF}; the second is \textbf{\themodel/feats--CF}, which employs graph convolutions on multimodal features instead of ID embeddings; the third is named as \textbf{CF+feats}, which does not consider latent item-item relationships and directly uses transformed multimodal features to replace the item representations learned from item graphs in Eq. (\ref{eq:cf}). Table \ref{tab:ablation} summarizes the performance and the relative improvements gained by \textbf{\themodel--CF} over \textbf{CF+feats}, from which we have the following observations:
\begin{itemize}
	\item Our method significantly and consistently outperforms original CF methods and other two variants with all three CF methods, which verifies the effectiveness of discovering latent structures and the flexibility of our plug-in paradigm. Specifically, we obtain 17.6\% average improvements over the \textbf{CF+feats} variants, which directly utilize multimodal features as side information of items.
	\item Based on the learned item graph structures, \textbf{\themodel/feats -- CF} employs graph convolutions on multimodal features. Our original method \textbf{\themodel--CF} utilizes the same learned structures but employ graph convolutions on item ID embeddings, which aims to directly model \emph{item affinities}. The improvements between two variants validate the effectiveness explicitly modeling item affinities. Multimodal features are used to bridge semantic relationships between items, which is important but not explicitly considered by existing methods.
\end{itemize}

\subsection{Sensitivity Analysis (RQ3)}
\label{section:rq3}

\begin{figure}
	\centering
	\subfloat[Varied \(k\) for Clothing]{
		\includegraphics[width=0.495\linewidth]{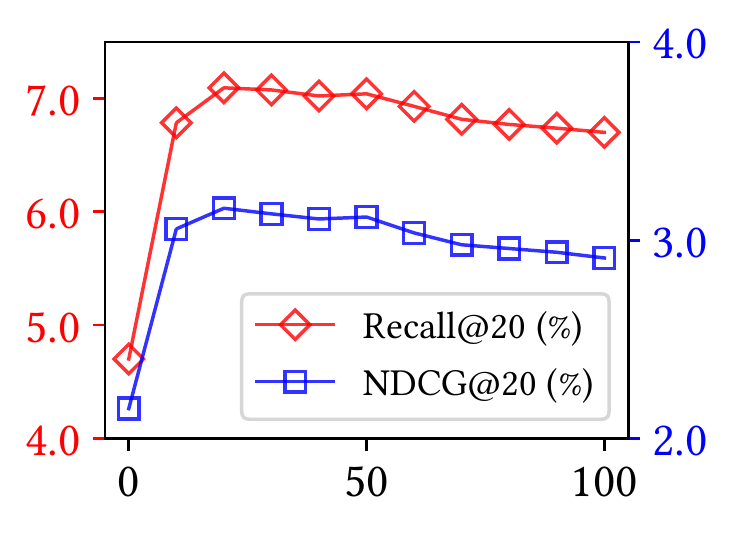}
	}\subfloat[Varied \(\lambda\) for Clothing]{
		\includegraphics[width=0.495\linewidth]{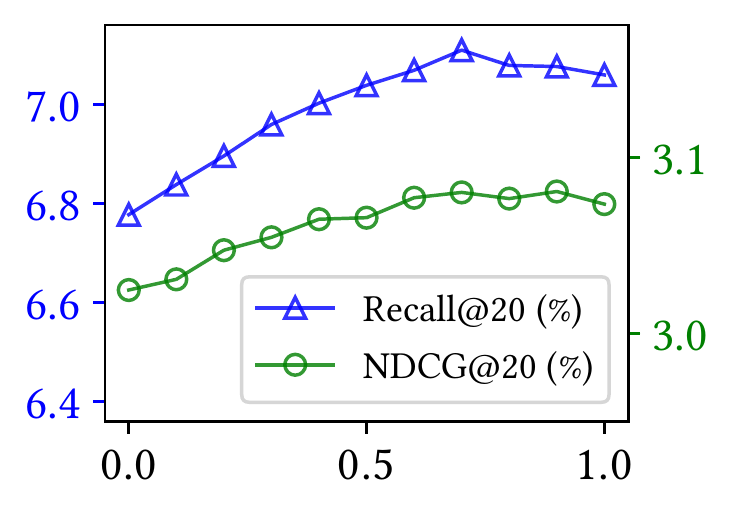}
	}

	\subfloat[Varied \(k\) for Sports]{
		\includegraphics[width=0.495\linewidth]{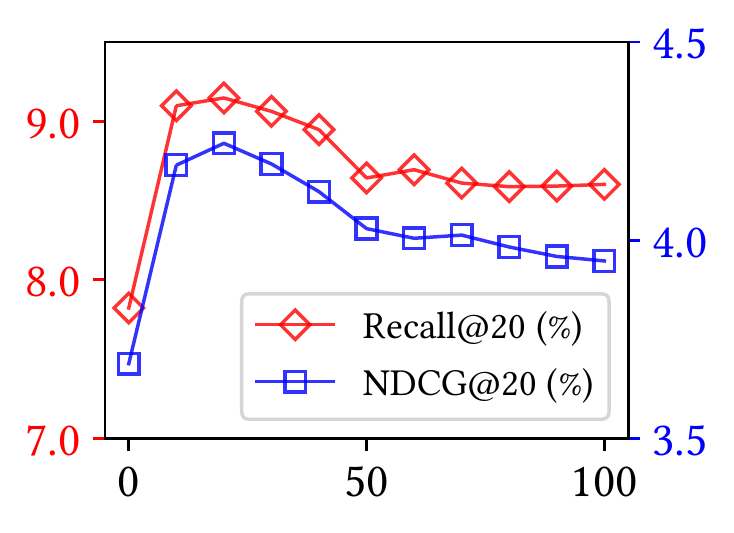}
	}\subfloat[Varied \(\lambda\) for Sports]{
		\includegraphics[width=0.495\linewidth]{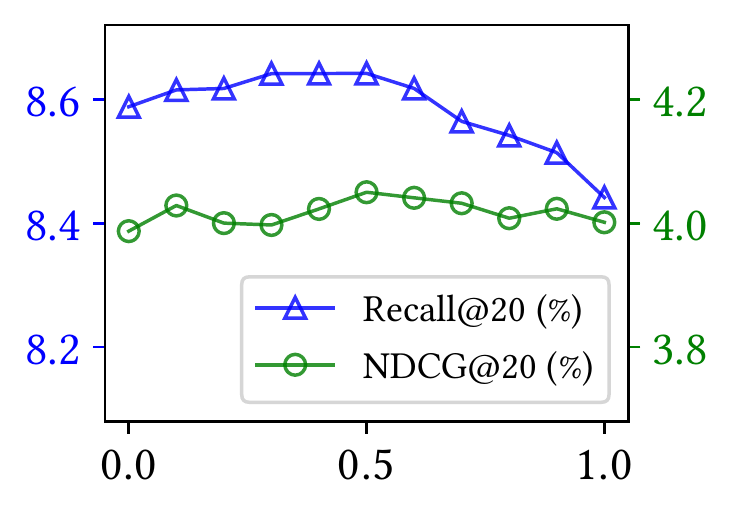}
	}

	\subfloat[Varied \(k\) for Baby]{
		\includegraphics[width=0.495\linewidth]{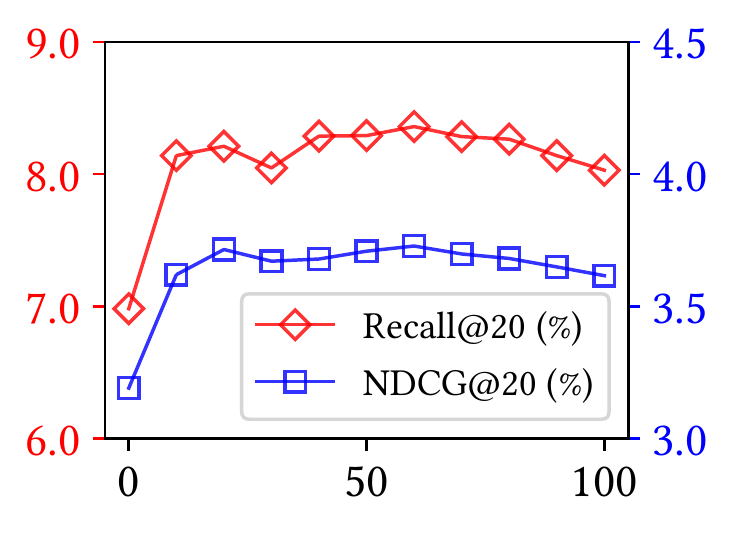}
	}
	\subfloat[Varied \(\lambda\) for Baby]{
		\includegraphics[width=0.495\linewidth]{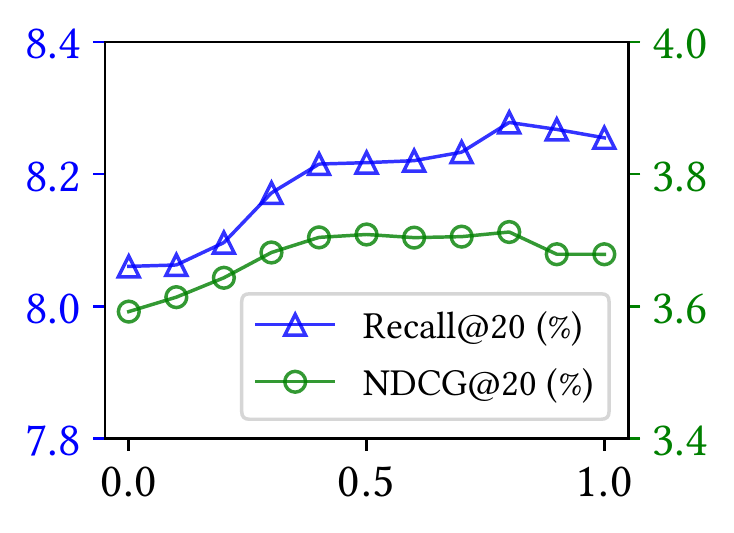}
	}
	\caption{Performances comparison over different hyperparameters settings.}
	\label{fig:hyperparams}
\end{figure}

Since the graph structure learning layer plays a pivotal role in our method, in this subsection, we conduct sensitive analysis with different hyperparameters on graph structure learning layers. Firstly, we investigate performance of \themodel-LightGCN with respect tor different $k$ value of the $k$-NN sparsification operation since $k$ value is important which determines the number of neighbors of each item, and controls the amount of information propagated from neighbors. Secondly, we discuss how the skip connection coefficient $\lambda$ affects the performance which controls the amount of information the from initial graph topology.

\subsubsection{Impact of varied $k$ values.}
Figure \ref{fig:hyperparams} reports the results of performance comparison. $k=0$ means no item relationships are included and the model is degenerated to original LightGCN. We have the following observations:
\begin{itemize}
	\item Our method gains significant improvement between $k=0$ and $k=10$, which validates the rationality of item relationships mined from multimodal features. Even if only a small part of the neighbors are included, we can obtain better item representations by aggregating meaningful and important information, which boost the recommendation performance.
	\item Furthermore, the performance first imporves as $k$ increases, which verifies the effectiveness of information aggregation along item-item graphs since more neighbors bring more meaningful information that helps to make more accurate recommendations.
	\item The trend, however, declines when $k$ continues to increase, since there may exist many unimportant neighbors that will inevitably introduce noisy to the information propagation. This demonstrates the necessity of conducting $k$NN sparsification on the learned dense graph.
\end{itemize}

\subsubsection{Impact of varied coefficients $\lambda$.}
Figure \ref{fig:hyperparams} reports the performance comparison. $\lambda=0$ means only consider the graph structure learned by the transformed high-level multimodal features, and $\lambda=1$ means we only consider the structure generated by the raw multimodal features. We have the following observations:
\begin{itemize}
	\item When we set $\lambda = 0$, the model obtain poor performance. It only learns graph structure from the transformed high-level features, completely updating the adjacency matrix every time, resulting in fluctuated training process.
	\item The performance first grows as $\lambda$ becomes larger, validating the importance of initial structures constructed by raw multimodal features. However, it begins to deteriorate when $\lambda$ continues to increase, since raw features are often noisy due to the inevitably error-prone data measurement or collection process.
	\item Overall, there are no apparent sharp rises and falls, indicating that our method is not that sensitive to the selection of $\lambda$. Notably, all models surpass the baselines (c.f. Table \ref{tab:experiments}), proving the effectiveness of item graphs.
\end{itemize}

\section{Related Work}
\subsection{Multimodal Recommendation}
Collaborative filtering (CF) has achieved great success in recommendation systems, which leverage users' feedbacks (such as clicks and purchases) to predict the preference of users and make recommendations.
However, CF-based methods suffer from sparse data with limited user-item interactions and rarely accessed items. To address the problem of data sparsity, it is important to exploit other information besides user-item interactions.
Multimodal recommendation systems consider massive multimedia content information of items, which have been successfully applied to many applications, such as e-commerce, instant video platforms, and social media platforms \cite{veit2015learning,McAuley:2015ip,He:2016dm,Cui:2021ks}.

For example, VBPR \cite{He:2016ww} extends matrix factorization by incorporating visual features extracted from product images to improve the performance.
DVBPR \cite{Kang:2017ds} attempts to jointly train the image representation as well as the parameters in a recommender model.
Sherlock \cite{he2016sherlock} incorporates categorical information for recommendation based on visual features.
DeepStyle \cite{Liu:2017ij} disentangles category information from visual representations for learning style features of items and sensing preferences of users.
ACF \cite{Chen:2017jj} introduces an item-level and component-level attention model for inferring the underlying users' preferences encoded in the implicit user feedbacks. 
VECF \cite{Chen:2019ga} models users' various attentions on different image regions and reviews.
MV-RNN \cite{cui2018mv} uses multimodal features for sequential recommendation in a recurrent framework.
Recently, Graph Neural Networks (GNNs) have been introduced into recommendation systems \cite{Wu:2019ke,Wang:2019er,zhang2020personalized} and especially multimodal recommendation systems \cite{Wei:2019hn,Wei:2020ko,li2020hierarchical}. MMGCN \cite{Wei:2019hn} constructs modal-specific graph and conduct graph convolutional operations, to capture the modal-specific user preference and distills the item representations simultaneously. In this way, the learned user representation can reflect the users’ specific interests on items. Following MMGCN, GRCN \cite{Wei:2020ko} focuses on adaptively refining the structure of interaction graph to discover and prune potential false-positive edges. 

The above methods directly utilize multimodal features as side information of each item. In our model, we step further by discovering fine-grained item-item relationships from multimodal features. 

\subsection{Deep Graph Structure Learning}
GNNs have shown great power on analyzing graph-structured data and have been widely employed for graph analytical tasks across a variety of domains, including node classification \cite{Kipf:2017tc,Zhu:2020vf,Zhu:2021gx}, link prediction \cite{Chen:2018vh}, information retrieval\cite{Zhang:2021vf, Yu:2021ka}, etc.
However, most GNN methods are highly sensitive to the quality of graph structures and usually require a perfect graph structure that are hard to construct in real-world applications \cite{Franceschi:2019uz}. Since GNNs recursively aggregate information from neighborhoods of one node to compute its node embedding, such an iterative mechanism has cascading effects --- small noise in a graph will be propagated to neighboring nodes, affecting the embeddings of many others. Additionally, there also exist many real-world applications where initial graph structures are not available. Recently, considerable literature has arisen around the central theme of Graph Structure Learning (GSL), which targets at jointly learning an optimized graph structure and corresponding representations. There are three categories of GSL methods: metric learning \cite{Chen:2020wu,Wang:2020bs,Li:2018wu}, probabilistic modeling \cite{Franceschi:2019uz,Zheng:2020tp,Luo:2021gg}, and direct optimization approaches \cite{Yang:2019fh,Jin:2020br,Gao:2020em}.

For example, IDGL \cite{Chen:2020wu} casts the graph learning problem into a similarity metric learning problem and leverage adaptive graph regularization for controlling the quality of the learned graph; DGM \cite{Kazi:2020vj} predicts a probabilistic graph, allowing a discrete graph to be sampled accordingly in order to be used in any graph convolutional operator. NeuralSparse \cite{Zheng:2020tp} considers the graph sparsification task by removing task-irrelevant edges. It utilizes a deep neural network to learn k-neighbor subgraphs by selecting at most k neighbors for each node in the graph.  We kindly refer to \cite{Zhu:2021ue} for a recent overview of approaches for graph structure learning.

In personalized recommendation, although user-item interactions can be formulated bipartite graph naturally, item-item relations remain rarely explored. To model item relationships explicitly, we employ metric learning approaches to represent edge weights as a distance measure between two end nodes, which fits for multimedia recommendation since rich content information can be included to measure the semantic relationship between two items. 

\section{Conclusion}
In this paper, we have proposed the latent structure mining method (\themodel) for multimodal recommendation, which leverages graph structure learning to discover latent item relationships underlying multimodal features. In particular, we first devise a modality-aware graph structure learning layer that learns item graph structures from multimodal features and fuses multimodal graphs. Along the learned graph structures, one item can receive informative high-order affinities from its neighbors by graph convolutions. Finally, we combine our model with downstream CF methods to make recommendations. Empirical results on three public datasets demonstrate the effectiveness of our proposed model.

\begin{acks}
This work was supported by National Key Research and Development Program (2018YFB1402600), National Natural Science Foundation of China (61772528, 62022083), Beijing National Natural Science Foundation (4182066), and Shandong Provincial Key Research and Development Program (2019JZZY010119).
\end{acks}

\bibliographystyle{ACM-Reference-Format}
\balance
\bibliography{acmmm2021}

\end{document}